\documentclass[prb,twocolumn,superscriptaddress,showpacs]{revtex4-2}

\usepackage{graphicx}
\usepackage{dcolumn}
\usepackage{bm}
\usepackage{hyperref}
\usepackage{url}
\usepackage{color}
\usepackage{amsmath}
\usepackage{siunitx}
\usepackage[running]{lineno}

\newcommand{\pcto}{PbCuTe$_{2}$O$_{6}$\,}

\begin{document}

\title{Crystal growth, characterization and phase transition of \pcto single crystals}

\author{A. R. N. Hanna}
\email[]{abanoub.hanna@helmholtz-berlin.de}
\affiliation{Institut f\"ur Festk\"orperphysik, Technische Universit\"at Berlin, Hardenbergstr. 36, 10623 Berlin, Germany}
\affiliation{Helmholtz-Zentrum Berlin f\"ur Materialien und Energie, Hahn-Meitner-Platz 1, 14109 Berlin, Germany}

\author{A. T. M. N. Islam}
\affiliation{Helmholtz-Zentrum Berlin f\"ur Materialien und Energie, Hahn-Meitner-Platz 1, 14109 Berlin, Germany}

\author{R. Feyerherm}
\affiliation{Helmholtz-Zentrum Berlin f\"ur Materialien und Energie, Hahn-Meitner-Platz 1, 14109 Berlin, Germany}

\author{K. Siemensmeyer}
\affiliation{Helmholtz-Zentrum Berlin f\"ur Materialien und Energie, Hahn-Meitner-Platz 1, 14109 Berlin, Germany}

\author{K. Karmakar}
\affiliation{Helmholtz-Zentrum Berlin f\"ur Materialien und Energie, Hahn-Meitner-Platz 1, 14109 Berlin, Germany}

\author{S. Chillal}
\email[]{shravani.chillal@helmholtz-berlin.de}
\affiliation{Helmholtz-Zentrum Berlin f\"ur Materialien und Energie, Hahn-Meitner-Platz 1, 14109 Berlin, Germany}

\author{B. Lake}
\affiliation{Helmholtz-Zentrum Berlin f\"ur Materialien und Energie, Hahn-Meitner-Platz 1, 14109 Berlin, Germany}
\affiliation{Institut f\"ur Festk\"orperphysik, Technische Universit\"at Berlin, Hardenbergstr. 36, 10623 Berlin, Germany}

\begin{abstract}
Single crystals of the three-dimensional frustrated magnet and spin liquid candidate compound \pcto, were grown using both the Travelling Solvent Floating Zone (TSFZ) and the Top-Seeded Solution Growth (TSSG) techniques. The growth conditions were optimized by investigating the thermal properties. The quality of the crystals was checked by polarized optical microscopy, X-ray Laue and X-ray powder diffraction, and compared to the polycrystalline samples. Excellent quality crystals were obtained by the TSSG method. Magnetic measurements of these crystals revealed a small anisotropy for different crystallographic directions in comparison with the previously reported data. The heat capacity of both single crystal and powder samples reveal a transition anomaly around \emph{1 K}. Curiously the position and magnitude of the transition are strongly dependent on the crystallite size and it is almost entirely absent for the smallest crystallites. A structural transition is suggested which accompanies the reported ferroelectric transition, and a scenario whereby it becomes energetically unfavourable in small crystallites is proposed.  
  
\end{abstract}


\date{\today}
\maketitle

\section{Introduction}

Strongly correlated systems have a wide range of fascinating properties that explain and verify the existence of novel quantum phenomena. Among these systems, some lattice geometries introduce unique examples of highly frustrated magnetic systems with novel behaviour such as quantum spin liquids~\cite{Balents2010,Fennell2009} and spin ice systems~\cite{Derzhko2015,Zhou2017,Morris2009, Benton2012}. The quantum spin liquid (QSL) is an unusual magnetic ground state caused by the interaction of quantum spins in certain magnetic materials with highly frustrated or competing interactions. QSLs are associated with the absence of long-range magnetic order and exhibit fractionalized excitations due to the quantum entanglement~\cite{Savary2016}.

\begin{figure}
\includegraphics[width=1 \columnwidth]{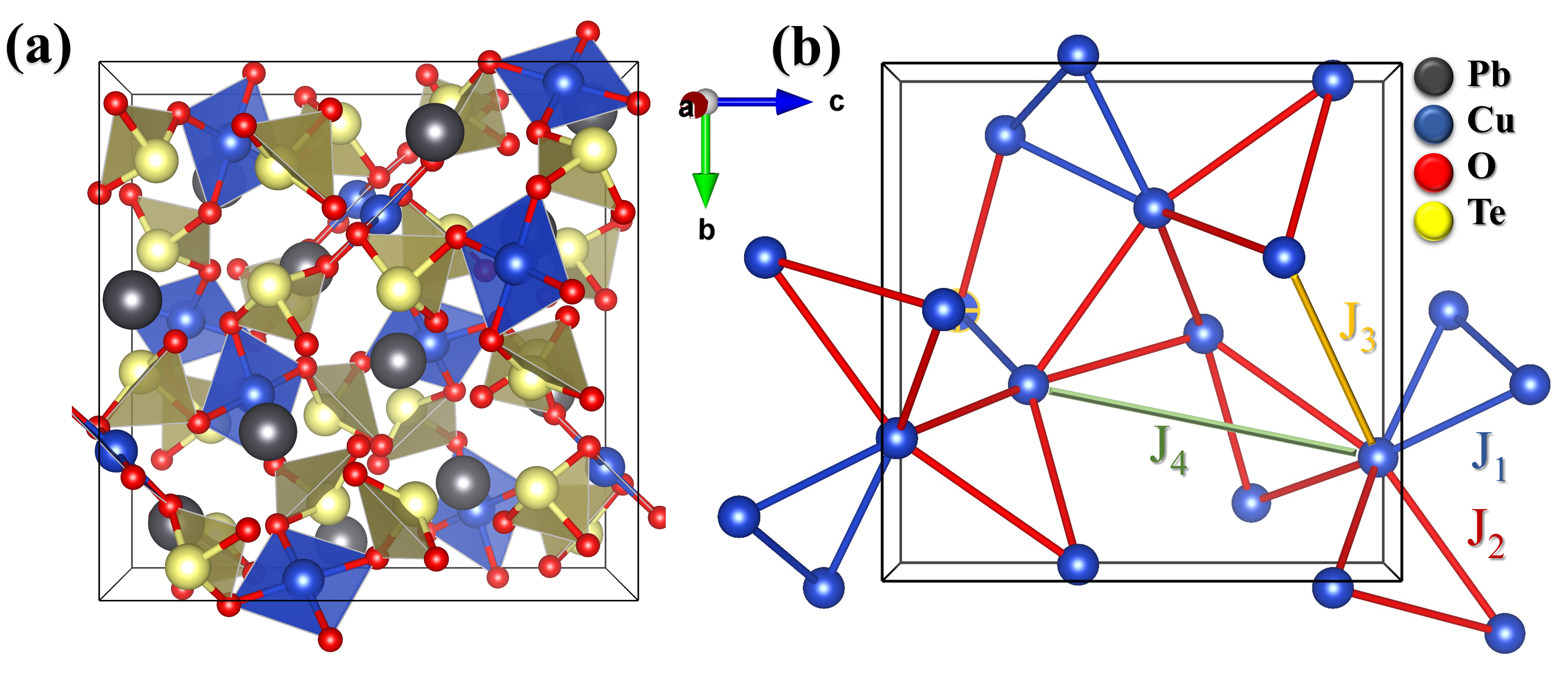}
\caption{(a) Visualization of the cubic structure of \pcto (Space group P4$_{1}$32) indicating the distribution of CuO$_4$ square plaquettes, TeO$_3$ units , and Pb atoms where the Pb, Cu, Te and O are colored grey, blue, yellow and red respectively. (b) The distribution of the Cu$^{2+}$ ions inside the lattice forming a 3-dimensional network of corner-shared triangles (hyper-hyperkagome lattice) where the first neighbor interactions (\emph{J$_1$}, blue lines) and the second neighbor interactions (\emph{J$_2$}, red lines) have similar strengths ~\cite{Koteswararao2014,Chillal2020}. While the weaker third interactions (\emph{J$_3$}, orange line) and fourth interactions (\emph{J$_4$}, green line) form the antiferromagnetic chains ~\cite{Chillal2020} }
\label{Fig1}
\end{figure}

\pcto (PCTO) is a rare example of a potential QSL hosted in a three-dimensional (3D) magnetic lattice known as the hyper-hyperkagome ~\cite{Chillal2020}.  This is a 3D network of corner shared triangles formed by the S=${1}/{2}$ Cu$^{2+}$ spins (three triangles per spin) similar to, but distinctly different from the more well-known hyperkagome lattice (two triangles per spin). The Cu$^{2+}$ spins in \pcto are coupled by various frustrated and non-frustrated antiferromagnetic exchange interactions~\cite{Koteswararao2014}. More detailed theoretical calculations suggest that the magnetic interactions extend to four nearest neighbours, where the first and second nearest neighbours ($J_1$,$J_2$) are equally strong and are involved in the frustrated hyper-hyperkagome lattice whereas the third and fourth neighbours (\emph{J$_3$, J$_4$})  form unfrustrated antiferromagnetic chains~\cite{Chillal2020}. Figure.~\ref{Fig1}(a) explains the crystal structure of \pcto where, the cubic structure (Space group P4$_1$32,  lattice constant =12.49~\AA\ at 300~\emph{K}) is formed from CuO$_4$ square plaquettes, TeO$_3$ units, and Pb atoms. The arrangement of magnetic Cu$^{2+}$ (S=$\frac{1}{2}$) ions forming the hyper-hyperkagome lattice is indicated in Fig.1(b) by the red and blue lines.

The strong frustration of the system is revealed by the absence of any static long-range magnetic ordering in the magnetic susceptibility, heat capacity, muon spin relaxation \emph{($\mu$SR)} and neutron diffraction measurements of the polycrystalline samples of PCTO down to 20    \emph{mK}~\cite{Koteswararao2014,Khuntia2016,Chillal2020}. While the heat capacity of the powder revealed a broad hump at $T\sim1$~\emph{K} and a weak anomaly at $T\sim$0.87~\emph{K} \cite{Koteswararao2014}, a long- range magnetic ordering was ruled out due to the absence of a $\lambda$-like anomaly. Inelastic neutron scattering of a polycrystalline sample at ($T=0.1$~\emph{K}) revealed a diffuse, non-dispersive continuum of magnetic excitations, a feature suggestive of the fractionalized excitations\cite{Chillal2020}. The diffuse scattering forms spheres in reciprocal space as seen in the single crystal samples~\cite{Chillal2020}, a 3D analogue of the 2D diffuse scattering rings observed in the kagome QSL Herbertsmithite and Ca$_{10}$Cr$_7$O$_{28}$~\cite{Han2012,Balz2016}, hence supporting the QSL interpretation of PCTO.

The crystal structure of PCTO and the experimental evidence for the absence of static magnetism have only been reported extensively in the polycrystalline samples. The information on structural and bulk properties of single crystals of PCTO remains inconclusive. For example, the sub-mm single crystals synthesised using the hydrothermal method report a larger lattice parameter(12.60~\AA, P4$_3$32)~\cite{weil2019, Wulff1997}. In addition, all single crystals also indicate a phase transition at $T\sim$1~\emph{K} \cite{Chillal2020,thurn2021}. Therefore, it is necessary to obtain large, high-quality \pcto single crystals suitable for the investigation of thermodynamic and magnetic properties as well as eventually inelastic neutron scattering measurements to study the possible QSL state.	

\pcto is a relatively new compound of which very little is known about its temperature-composition phase equilibrium. The differential thermal analysis elucidated that \pcto has a high tendency toward incongruent melting; our initial experiments revealed a decomposition at \emph{600$^\circ$C} and the presence of the other phases Cu$_3$TeO$_6$, Pb$_2$Te$_3$O$_8$ which are clear indications of the instability of the melt. Besides, due to its proximity to the crystalline phase PbCuTe$_{2}$O$_{7}$ ~\cite{Yeon2011} which has a slightly higher oxidation state, the growth atmosphere of \pcto needs to be optimized carefully. For these reasons, the growth of large and high-quality single crystals of \pcto is considered quite challenging. Among different crystal growth techniques, the Travelling Solvent Floating Zone technique (TSFZ) is known for the successful growth of many such complex oxide compounds~\cite{Koohpayeh2008,Islam2011}. It provides high purity crystals because of its advantage in controlling the growth atmosphere and the absence of contact of the melt with any other material (e.g. Crucible, wires, covers, etc). Many novel quantum magnets were grown using this technique such as the quantum spin liquid candidate Ca$_{10}$Cr$_7$O$_{28}$~\cite{Balz2016}, CaCr$_2$O$_4$  with distorted triangular lattice~\cite{Toth2011} and the candidate quantum spin-ice system Pr$_2$Hf$_2$O$_7$~\cite{Anand2016}. However, some compounds have very low surface tension in the molten form and a stable melt zone is extremely difficult to maintain in a floating zone growth, as sudden flow-down of the melt can cause frequent disruption of growth. For melts of such nature, it is more advantageous to contain the melt in a crucible and grow the crystal by a pulling method such as Top-Seeded Solution Growth (TSSG)~\cite{Polgar1997,Dagdale2016}. 

Here we report the growth of large single crystals of \pcto by both the TSFZ technique in an optical floating zone furnace and TSSG in the high frequency induction furnace following the Czochralski (CZ) method. Single crystals grown by these two methods were characterized by powder x-ray diffraction, Laue diffraction, polarized optical microscopy and magnetic susceptibility, and the effects of different growth techniques on the quality of the single crystals are discussed. Finally by performing heat capacity measurements down to low temperatures we found a clear Lambda anomaly at $T\sim1$ \emph{K} in the single crystals grown by both methods, in contrast to the results for polycrystalline samples. This feature was first discovered and reported in the recent paper, C. Thurn {\it et. al.}~\cite{thurn2021}, where it was assigned to a ferroelectric transition   accompanied by strong anisotropic distortions by its signatures in a combination of thermal expansion, specific heat, dielectric and ac susceptibility measurements on the crystals that we describe here. In this current paper, we perform a detailed investigation into the effects of crystallite size, quality and stoichiometry and show that the transition is an intrinsic property unrelated to impurities or defects, which however is suppressed for sufficiently small crystallite size. The origins of this transition and its consequences for the magnetic properties of \pcto are discussed.

\section{Samples \& Experimental Methods}
\subsection{Travelling Solvent Floating Zone (TSFZ) growth}
For crystal growth, first a polycrystalline powder of \pcto  was synthesized by solid-state reaction of high purity powder of PbO (99.99\%, Alfa Aesar), CuO (99.995\%, Alfa Aesar) and TeO$_2$ (99.9995\%, Alfa Aesar) mixed thoroughly in the 1:1:2 molar ratio in ethanol. After mixing, the stoichiometric composition was calcined in a platinum crucible under flowing argon atmosphere in a tube furnace for 12 hours at 540$^\circ$C twice with an intermediate grinding. The powder was then packed into rubber tubes and compressed in a cold hydrostatic press at 2000 bar to form uniform cylindrical rods of length 70-80 mm and diameter around 6 mm. The pressed rods were then hanged in the tube furnace and sintered in flowing argon for 12 hours at 550$^\circ$C to form dense rods, ready to be used as the precursor for crystal growth.

Crystal growth was first carried out in an optical floating zone furnace (Crystal Systems Corp., FZ-T 10000-H-VI-VPO) equipped with four 150 W Tungsten halide lamps focused by ellipsoidal mirrors. The growths were performed under different atmospheres from 0.1-0.2 MPa Argon flow pressure and growth rates ranging from 0.5 to 2.0~mm/h. For crystal growth by the TSFZ technique, we needed to find an optimum solvent, which stays in equilibrium with \pcto  when melted. Since no phase equilibrium diagram of PCTO exists, we carefully investigated the reported binary phase diagrams of PbO-TeO$_2$, CuO-TeO$_2$, and CuO-PbO to get  insights into the possible phase relations in the compound ~\cite{Robertson1976,Stavrakeva1990,Torres2008}. Therefore, we constructed the ternary phase diagram of CuO-PbO-TeO$_2$ as in Fig. 2(a) where the \pcto phase is shown along with the related compounds and their melting temperatures. The blue lines represent the mole fraction of each component of stoichiometric \pcto. In addition, thermogravimetry and differential thermal analysis (TG-DTA) were performed on the pure polycrystalline powder of PCTO (NETZSCH TG 209 F3 Taurus) with a heating rate of 5$^\circ$ C/minute over the range 425-650 $^\circ$ C under Argon flow of 10~cc/minute. From the TG-DTA on PCTO, we observe several interesting thermal events (Fig. 2(b)). We expect that the broad exothermic peak at 556$^\circ$ C corresponds to the decomposition of PCTO, while the endothermic peak at  561$^\circ$ C corresponds to the melting of the compound. X-ray powder diffraction shows that the phases after heating to  670$^\circ$C were \pcto ,Pb$_2$Te$_3$O$_8$ and Cu$_3$TeO$_6$ providing further confirmation of the decomposition of PCTO.

\begin{figure}
\includegraphics[width=1 \columnwidth]{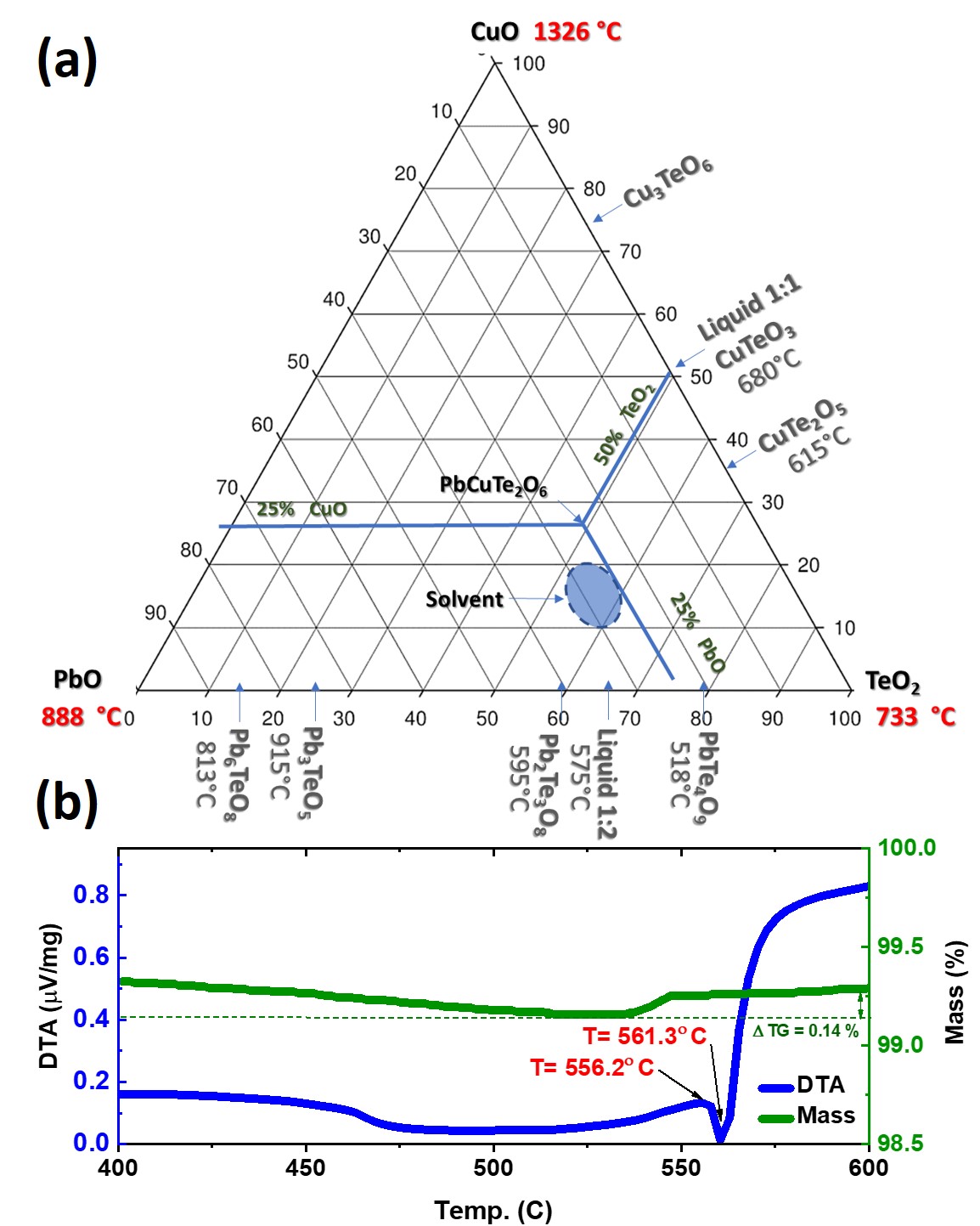}
\caption{(a) Ternary Phase diagram of \pcto precursors (PbO,CuO and TeO$_2$). The blue lines represent mole fraction of  stoichiometric \pcto. The blue arrows on the axis represent the stable phases and their melting temperatures.  The  point of intersection of the blue lines represents the area at which \pcto is expected to form. The dotted blue oval represents the solvent composition for TSFZ gowth . (b) Thermogram of \pcto  powder showing TGA (green line)and DTA (blue line). The broad peak in DTA between 460-550$^\circ$\emph{C} might be attributed to the melting of small crystallites }
\label{Fig2}
\end{figure}

Looking into the three binary phase diagrams we conclude that the most probable composition of the solvent with a melting temperature below the PCTO decomposition temperature (556$^\circ$\emph{C}) can be found within the blue oval shape shown in Fig. 2(a).  For this reason, during the initial experiments we have used a solvent of composition with excess PbO , TeO$_2$. The solvent compound was prepared in the same way as the stoichiometric PCTO by solid-state reactions. Approximately 0.5 grams of the solvent was melted and attached to the tip of the stoichiometric feed rod. To begin the crystal growth , only the solvent tip of the rod was re-melted and was attached to the seed crystal to form a molten-zone.

The differential thermal analysis technique (DTA) combined with the thermogravimetric analysis  (TGA) is generally used to identify any thermal event related to the sample such as melting, crystallization, decomposition and  structural transition. The melting of any compound is characterized by an endothermal peak in DTA without change in the mass. Although \pcto has an incongruent melting nature, the TG-DTA measurement together with our initial growth experience showed that the incongruent melting temperature of PCTO is very close to the peritectic point and no mass loss occured due to evaporation of any other phase such as PbO.    So, it should be possible to grow the PCTO compound without using a non-stoichiometric solvent. For this reason, subsequent growths were carried out just by melting the tip of the stoichiometric feed rod, without attaching any off-stoichiometric solvent to it. The growth rate was kept at 1 mm/hour allowing the melt composition to change slowly in the next few millimetres to the desired solvent composition that would be in equilibrium with PCTO phase. 

\subsection{Top-Seeded Solution Growth (TSSG)}
Top-seeded solution growth of PCTO was carried out in a Czochralski furnace (Mini Czochralski Oxypuller 05-03 Cyberstar, France) equipped with an RF heating coil and a highly sensitive weighing device attached to the pulling rod to measure growth mass in real time. From our experience of TSFZ growth, we found that it was possible to grow PCTO without adding an off-stoichiometric solvent if growth is done at a slow rate. So, a stoichiometric polycrystalline powder of PCTO was prepared as before by solid-state reaction in argon flow. A platinum crucible of 40 mm  diameter and 20 mm  height was filled with 40 grams of well-reacted powder and positioned at the centre of the upper part of the coil using a suitable ceramic heating setup. A platinum rod shaped like a cylinder of 3 mm  diameter and 20 mm length was attached to the pulling shaft to be used as a seed. To minimize the contamination of the chamber, the seed and the powder in the crucible were enclosed in a quartz tube covered by a ceramic lid with a hole to allow the seed movement. The growth chamber was purged and filled with high purity Argon gas a couple of times before heating the coils. 
\begin{figure}
\includegraphics[width=1 \columnwidth]{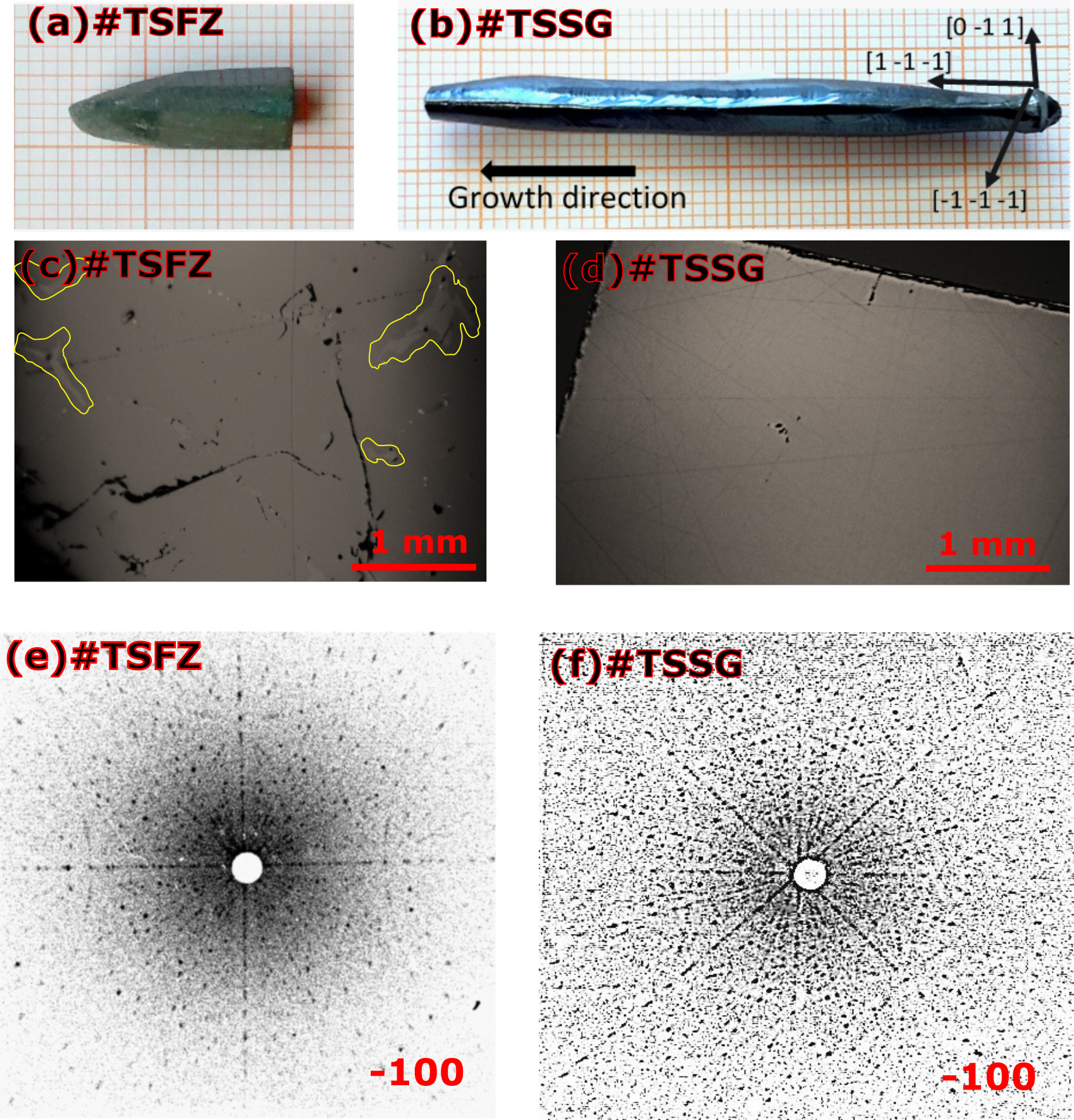}
\caption{(a) Representative single crystal of \pcto grown by optical floating zone method (Crystal TSFZ). (b) crystal grown by Top -seeded solution growth method (Crystal TSSG). (c,d) Optical  images of the crystal cross-section indicating the quality of  the  TSFZ crystal and  TSSG crystal respectively. The yellow outlines indicate the presence of the impurity phase (Pb$_2$Te$_3$O$_8$) in the TSFZ crystal. (e ,f) Typical X-ray Laue backscattering photographs along the direction [100] of TSFZ and the TSSG crystal.} 
\label{Fig3}
\end{figure}
The growth was performed under an ambient argon atmosphere with a flow rate of 10~cc/minute.  The power of the radio frequency (RF) heater was increased slowly to 6.5\% over 3 hours to prevent any evaporation of the precursor. After complete melting, the platinum seed was dipped into the melt and rotated at 5 rpm for mixing. The platinum rod was then pulled upwards at a rate of 0.7 mm/h to initiate the growth from its tip. The growth rate in milligrams/hour was monitored by the highly sensitive weighing device attached to the pulling rod. During the growth, as the amount of the melt in the crucible decreased with time, the diameter of the as-grown crystal was controlled by manipulating the growth speed and the power to the RF coil based on visual observation of the growth and the data obtained from the weighing device. The power of the RF heater was decreased slowly from 6.5 to 6.0 \% and the pulling speed was increased from 0.7 to 1.0 mm/h to maintain a stable growth with a mostly constant growth rate of 300 milligram/hour. The total growth duration was about 72 hours  resulting in a cylindrical rod of length 65 mm and diameter 4 mm. 

The single crystallinity of the crystals grown by both the TSFZ and TSSG methods was checked by X-ray Laue diffraction and polarized optical microscopy (see Fig.3). Pieces from each crystal were ground and checked with powder x-ray diffraction (PXRD) (Brucker D8) for phase purity. The direct current (DC) magnetic susceptibility $\chi$ versus temperature T and isothermal magnetization M versus magnetic field H measurements were performed using a superconducting quantum interference device (SQUID-VSM, Quantum Design, Inc.). The temperature dependence of the heat capacity was measured using a physical property measurement system (PPMS, Quantum Design, Inc.),Measurements down to  $T=0.35$   \emph{K} were achieved using a $^{3}$He insert. All the growths and characterization measurements were performed at the Core Lab for Quantum Materials, Helmholtz Zentrum Berlin (HZB). 

\section{Results and Discussion}
\subsection{Travelling  Solvent Floating Zone growth}
A Single crystal grown by the TSFZ method is shown in Fig. 3(a). The molten zone of growth was observed to be quite unstable. Continuous monitoring was required to maintain the growth, because of the low surface tension of the melt, it has a tendency to flow down. This could be attributed to  the presence of the heavy metallic elements Pb and Te. The optimum conditions for our growth were found to be at a speed of 1 mm/hour in an atmosphere of 0.2 MPa. Growth was done from a stoichiometric feed rod, as the use of a solvent did not seem to have a significant advantage from our experience of previous growths.

The cross-section in Fig.3(c) indicates inclusions (outlined in yellow) in the bulk indicating small amounts of impurities in the single crystal. Nevertheless, the Laue diffraction taken at several points shows clear patterns belonging to the P4$_1$32 symmetry of PCTO. Fig.3(e) shows the Laue image with incident x-rays along the -100 direction.

\subsection{Top-Seeded Solution Growth} 
Figure. 3(b) shows an as-grown crystal grown by the TSSG method of dimensions of length 65 mm and diameter 4 mm. Unlike the crystal grown by the TSFZ method,  TSSG crystal shows nicely developed facets and a shiny surface with a metallic lustre. We believe that the different heating technique as well as the confinement of the powder in a crucible allowed better melting conditions and prevented the melt from flowing downwards as occurred in the TSFZ growth. No inclusion of any impurity phase was found in the cross-section of the single-crystal as observed by polarized optical microscopy (Fig. 3(d)) or identified by X-ray powder diffraction of crushed crystal from this growth. 

X-ray Laue backscattered images taken from both crystals are shown in Fig. 3(e and f). The TSSG crystal shows considerably sharper Laue photographs, a further testimony of the superior quality of the crystal.  

\begin{table}
\centering
\begin{tabular}{|c|c|c|c|} 
\hline
\bf Parameter & \bf TSSG  & \bf  TSFZ & \bf	Poly  \\ 
\hline
Lattice parameters &	12.49664(24)	& 12.50196(15)	& 12.49874(19)\\
\hline
Vcell (\AA$^3$) & 	1951.5613	& 1954.043	& 1952.534 \\
\hline
GOF  &	2.3             & 0.32          & 1.7 \\
\hline
Rp (\%)  & 	2.47	& 3.19	& 4.86  \\
\hline
Rwp \%)	 & 3.55	 & 4.28	& 6.30 \\
\hline
Rexp(\%) &	1.51	& 12.89 &	3.46  \\
\hline
\end{tabular}
\caption{The Rietveld refinement parameters of \pcto samples including the cell volume (Vcell), goodness of fit (GOF) and the R-Factors: profile R-Factor (Rp),weighted-profile R-factor (Rwp) and expected R-Factor (Rexp).  The (Pb$_2$Te$_3$O$_8$) phase present in the TSFZ sample was considered.}
\label{tab1}
\end{table}
\begin{figure}
\includegraphics[width=1 \columnwidth]{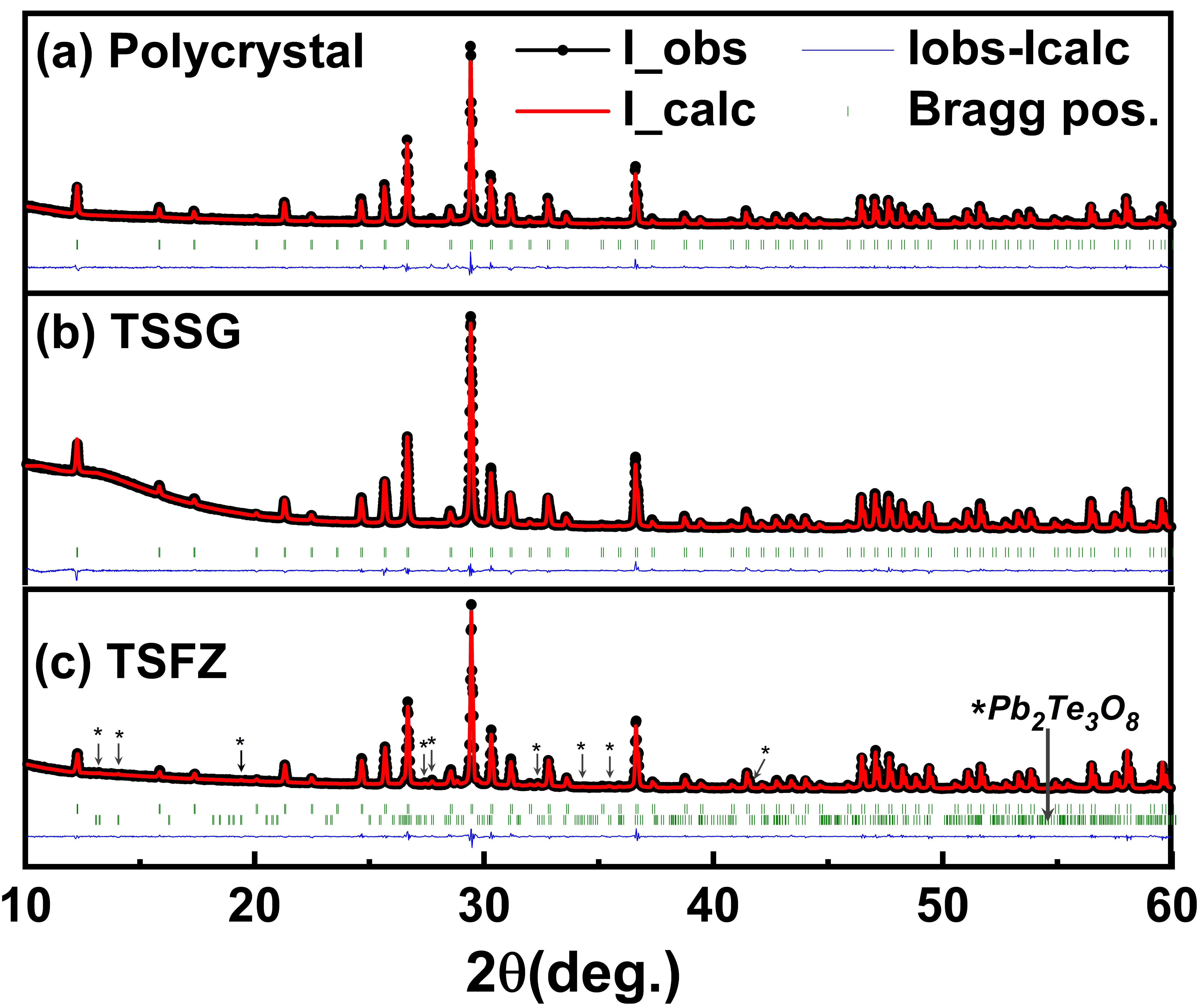}
\caption{Comparison of the Rietveld refinement for the powder x-ray diffraction of \pcto crushed samples (a) single crystal TSSG, (b)single crystal TSFZ(with Pb$_2$Te$_3$O$_8$ additional phase), (c)polycrystalline powder. The solid red line through the experimental points is the Rietveld refinement profile  for the \pcto-type cubic structure (space group P4$_1$32). The short green vertical bars mark the Bragg peaks positions and the lowermost blue curve represents the difference between the experimental and calculated intensities. The refinement parameters are summarized in Table.~\ref{tab2}. The pronounced background in TSSG sample is an instrumental error. }

\label{Fig4}
\end{figure}

\subsection{Crystallography}
 
The diffraction patterns of polycrystalline as well as a crushed powder from the crystals of PCTO have been measured using x-rays ($\lambda$=1.54 \AA) at room temperature. The patterns were refined by the Rietveld method using the FullProf software~\cite{Carvajal1993}. In all the three samples, the best refinements of the diffraction patterns were obtained by considering a cubic structure of space group P4$_1$32 as shown in Fig.4. The obtained lattice parameters and the quality parameters of the refinement are listed in Table. ~\ref{tab1}. The lattice parameters are 12.498 \AA, 12.4967 \AA, 12.502 \AA  ~for polycrystalline (Rwp=6.30), TSSG (Rwp=3.55) and TSFZ (Rwp=4.28) crystals respectively. These values agree very well with the previously reported polycrystalline \pcto and confirm that the space group is P4$_1$32~\cite{Koteswararao2014,Chillal2020}. Allowing the occupancy of the copper ion to vary showed that the refinement favored the stoichiometric sample and did not reveal any disorder of magnetic ion in  the \pcto system. A complete list of refined parameters and possible impurities in the samples are given in the supplementary data. 

Furthermore, the TSSG sample does not show any additional Bragg peaks that do not belong to PCTO, reflecting the single-phase nature of these samples (see fig. 4(b)). However, the TSFZ crystal contains some additional Bragg peaks related to the nonmagnetic phase of composition Pb$_2$Te$_3$O$_8$ (space group: Amam) as shown in fig.4(c) by the stars and lower set of green bars,and Fig. 3(c) indicates the Pb$_2$Te$_3$O$_8$ impurity in the microscopy image. This phase has been included in the refinement revealing a phase fraction of ~6\%. However, we do not see evidence for Oxygen deficiency in the \pcto, which if present, is expected to be less than 0.5\%. These values are supported by TGA measurements of different samples of \pcto in oxygen atmosphere which revealed a close to stoichiometric oxygen content (see supplementary information).  

In a less stable TSFZ growth, 3-4\% of Cu$_3$TeO$_6$ was also observed in addition to the Pb$_2$Te$_3$O$_8$ impurity. The presence of both Pb$_2$Te$_3$O$_8$  and Cu$_3$TeO$_6$ phases in the TSFZ crystal indicated the presence of a eutectic mixture in the growth temperature range. 

The microscopic differences in the PXRD patterns of the single crystals TSSG, TSFZ, and polycrystal \pcto are  presented in figure 4 along with the individual  Rietveld refinements to the P4$_1$32 structure of PCTO. The fit parameters corresponding to the refinement are listed in Table.~\ref{tab1}. Here, it is clear that both the polycrystal and  TSSG have comparable values of the refinement quality, whereas  the TSFZ has a different values due to the presence of the impurity phase. The presence of the  impurity phase in the TSFZ crystal is highlighted and indicated by the additional set of Bragg positions.Finally we would like to point out that these patterns can also be refined with similar quality by considering the P4$_3$32 space group which is the enantiomorphic pair of P4$_1$32 used in our analysis.

\subsection{Magnetic properties}

\begin{figure}
\includegraphics[width=1 \columnwidth]{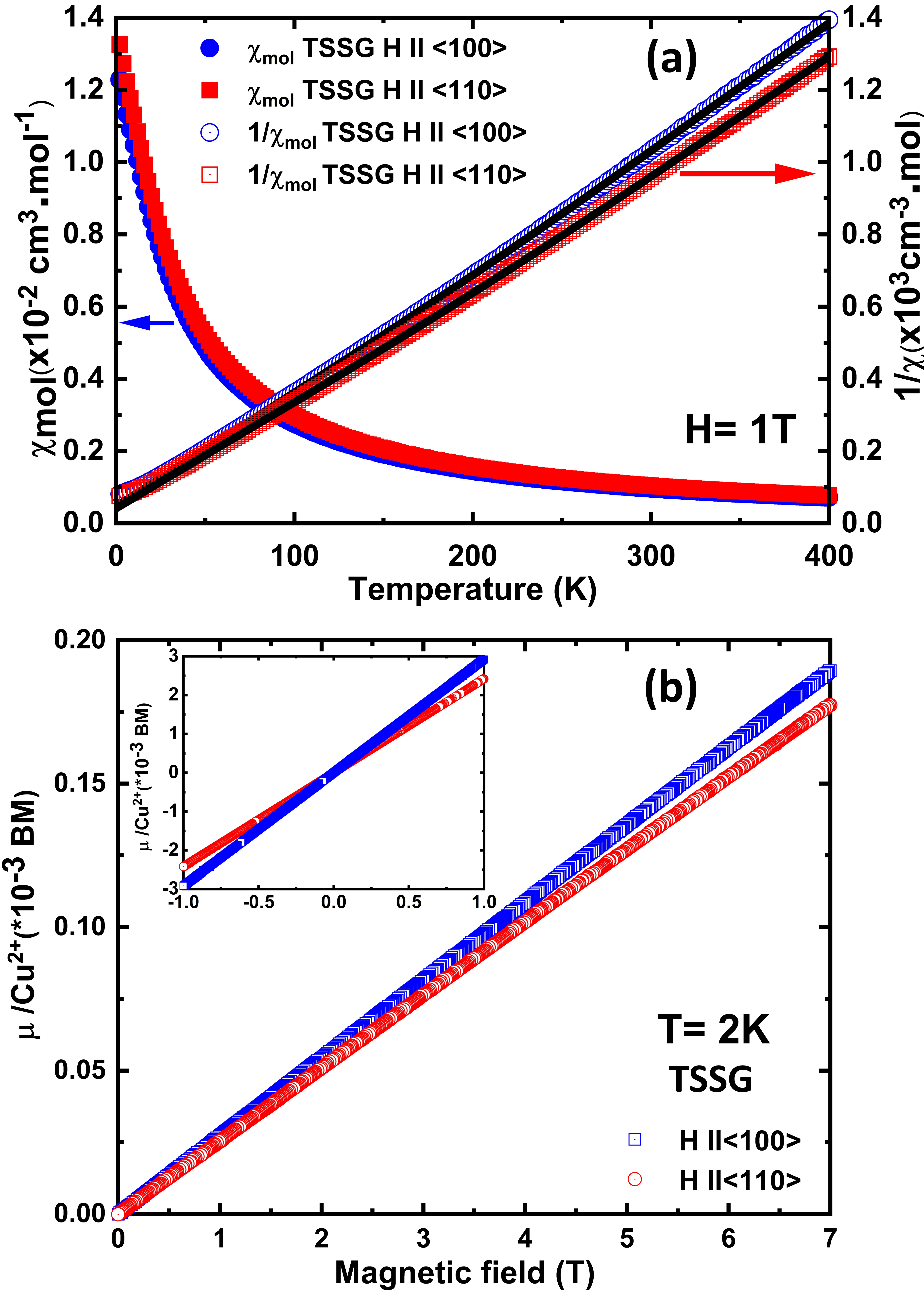}
\caption{(a) The temperature dependence of the magnetic susceptibility $\chi$(T) (left y-axis) and  1/($\chi$) data (right y-axis) of \pcto TSSG crystal with magnetic field applied along the [100] and [110] crystallographic directions in zero-field-cooled (ZFC) condition with a magnetic field of 1Tesla. The inverse susceptibility data is fitted to the inverse-Curie-Weiss law.(b)Field dependence of the induced magnetic moment along [110], [100] directions for the TSSG crystal at 2 \emph{K}. The inset shows the hysteresis loop at 2 \emph{K} up to a field range of 1 Tesla.}
\label{Fig5}
\end{figure}

Figure.~\ref{Fig5}(a) elucidates the magnetic measurements performed on the two principle directions of TSSG sample of \pcto. The temperature dependence of the molar DC magnetic susceptibility was measured in the temperature range $2-400$~\emph{K}, using three different applied magnetic fields of 0.1~T, 0.5~T and 1~T. As plotted in figure 5(a) for 1~T, the susceptibility revealed a slight anisotropy between $[100]$ and $[110]$ directions at high temperatures in all the measured fields. To quantify these results, the inverse susceptibility data were fitted by the modified Curie-Weiss law considering the total susceptibility, $\chi=\chi_{0}+c/(T-\theta_{CW})$ where c is the Curie constant, $\theta_{cw}$ is the Curie-Weiss temperature and $\chi_{0}$ is a sum of temperature independent contributions from the core diamagnetism $\chi_{core}=-1.65\times10^{-4}$~mol/cm$^3$ and the Van-Vleck paramagnetic susceptibility $\chi_{vv}$. 

The earlier measurements of inelastic neutron scattering of the single crystal and polycrystalline samples indicated that the magnetic excitation spectrum for PCTO shows only broad excitations extending up to 4~\emph{meV}. This suggests that on the temperature scale, the  magnetic correlations could be present up to $\sim$45~\emph{K}. To get a good-quality fitting for paramagnetic susceptibility, we therefore varied the lower bound of the fit range above $T_l=60$~\emph{K}. The upper bound of the fit was fixed to 300~\emph{K}~ for these samples to be compared with the previously reported data on polycrystalline and TSFZ samples \cite{Koteswararao2014,Chillal2020}. For the TSFZ sample, the non-magnetic impurity Pb$_2$Te$_3$O$_8$ was taken into consideration by normalizing the mass accordingly.  The fits were found to be stable for \emph{$T_l$} up to 130~\emph{K} and describe the full range of the linear inverse susceptibility. The resulting fit parameters summarized in Table.~\ref{tab2} for 1~T data, suggest a uniform $\theta_{cw}\sim-19$ \emph{K} for both the directions, very close to the previous reports. However, we observe a slightly different values of $g-$ factor calculated from the fits, namely $g_{[110]}\sim$2 and $g_{[100]}\sim$1.9 indicating that PCTO has a slight deviation from Heisenberg behaviour.  Possible reasons for the anisotropy could be the Dzyaloshinskii–Moriya interaction (DMI) which is allowed since the space group is non-centrosymmetric. Another reason could be a single ion ansiotropy of Cu$^{2+}$ due to unquenched orbital momentum. Additionally, anisotropic $g-$factors have also been previously observed in systems where Cu$^{2+}$ sits at the centre of plaquettes formed by oxygen ions that are oriented in a preferential direction with respect to crystal axes~\cite{Zvyagin2006,Katia2016}. In \pcto too, these plaquettes lie in the 6 different local $[110]$ planes resulting in a weaker net anisotropy. The effects of weak anisotropy are also visible in the  measurements of the isothermal magnetization (see fig.~\ref{Fig5}(b))  where the effective moment per Cu ion along the two directions of the single crystal differ as the field increases. 

\begin{table}
\centering
\begin{tabular}{|c|c|c|c|c|c|c|} 
\hline
\bf Sample & 
\bf H $||$ & 
 
\bf	\begin{tabular}[c]{@{}l@{}} range \\   (K) \end{tabular}   & 
\bf \begin{tabular}[c]{@{}l@{}@{}}	$\chi_{vv}$\\(mol\\/cm$^3$)	\end{tabular}  & 
\bf \begin{tabular}[c]{@{}l@{}} $\Theta_{CW}$\\(K) \end{tabular} &
\bf	\begin{tabular}[c]{@{}l@{}@{}}  $c$ \\(emu.K\\/mol) \end{tabular}    & 
\bf	\begin{tabular}[c]{@{}l@{}@{}}$g$-factor \\ \end{tabular}   \\
\hline
TSSG &	100	&  \begin{tabular}[c]{@{}l@{}}100-\\300	\end{tabular}	& \begin{tabular} [c]{@{}l@{}} 7.0 (1) \\ *10$^{-5}$ \end{tabular} &
\begin{tabular}[c]{@{}l@{}} -18.9 \\$\pm$0.3 \end{tabular}	&  0.34 & 	1.98$\pm$0.1 \\
\hline
TSSG &	110	&  \begin{tabular}[c]{@{}l@{}}100-\\300	\end{tabular}	& \begin{tabular} [c]{@{}l@{}} 5.6(2) \\ *10$^{-5}$ \end{tabular} &
\begin{tabular}[c]{@{}l@{}}  -18.6 \\${\pm}$0.3 \end{tabular}	& 0.37	& 	1.90$\pm$0.1 \\
\hline
TSFZ &	110	&  \begin{tabular}[c]{@{}l@{}}100-\\300	\end{tabular}	& \begin{tabular} [c]{@{}l@{}} 8.19 (2) \\ *10$^{-5}$ \end{tabular} &
 \begin{tabular}[c]{@{}l@{}} -18.4 \\${\pm}$  0.2 \end{tabular}	& 0.43	& 	2.14$\pm$0.1 \\
\hline
Poly &	 	& \begin{tabular}[c]{@{}l@{}}100-\\300	\end{tabular}	& \begin{tabular} [c]{@{}l@{}}6.7(2) \\ *10$^{-5}$ \end{tabular} &
 \begin{tabular}[c]{@{}l@{}} -20.2 \\${\pm}$  0.2 \end{tabular}	& 0.41	& 	2.09$\pm$0.1 \\
\hline
\begin{tabular}[c]{@{}l@{}} Reported\\~\cite{Koteswararao2014} \end{tabular} &	 	&  \begin{tabular}[c]{@{}l@{}}20-\\300	\end{tabular}	& \begin{tabular} [c]{@{}l@{}}4.5  \\ x10$^{-5}$ \end{tabular} & \begin{tabular}[c]{@{}l@{}} -22.0 \\${\pm}$  0.5 \end{tabular}	& \begin{tabular} [c]{@{}l@{}}0.38   \end{tabular}	& 	2.0 \\
\hline
\end{tabular}
\caption{ Curie–Weiss fitting parameters obtained from the analysis of magnetic susceptibility of \pcto at a field of 1~T.}
\label{tab2}
\end{table}

\begin{figure}
\includegraphics[width=1 \columnwidth]{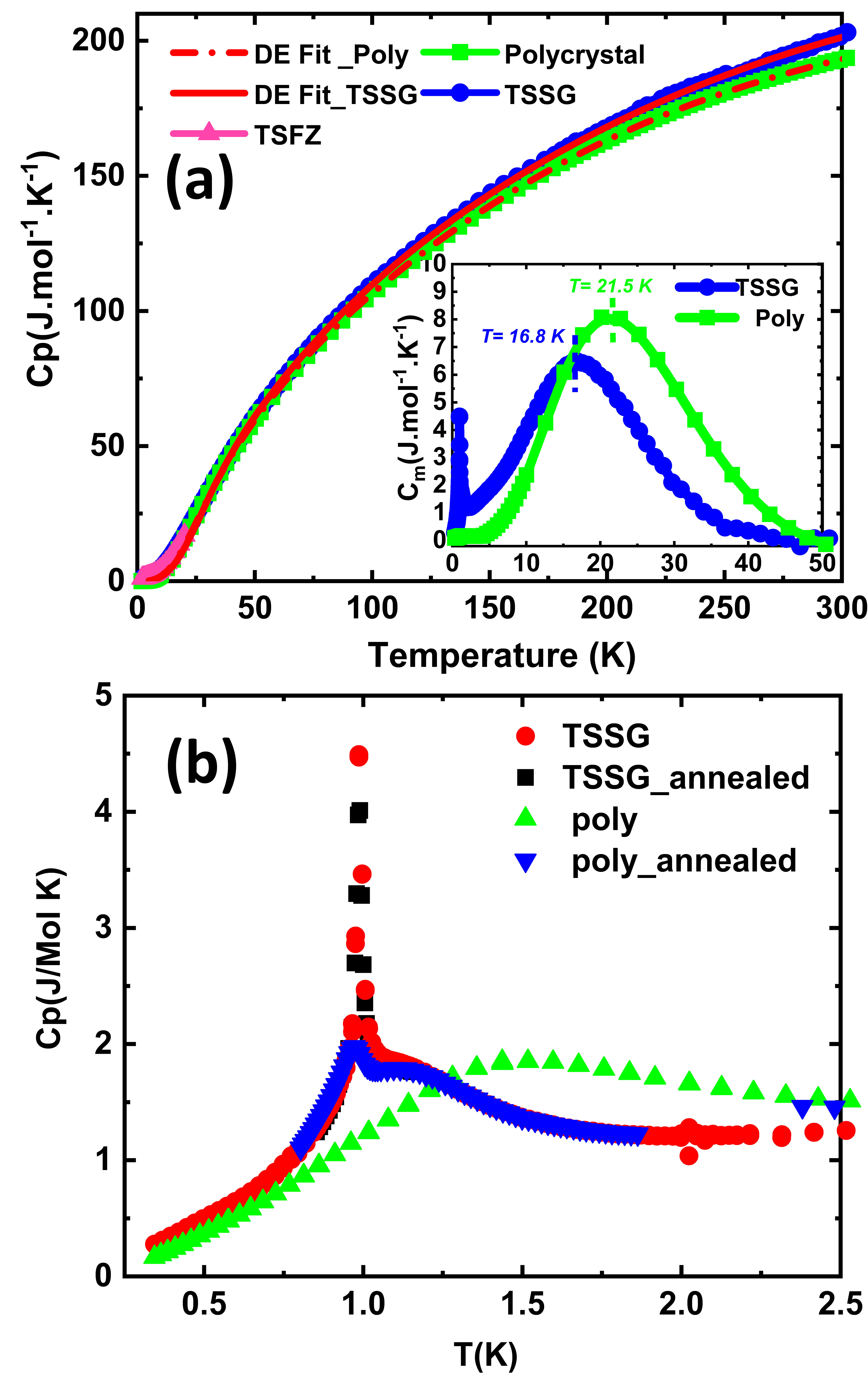}
 
\caption{Temperature dependence of the heat capacity of \pcto in the temperature range 0.3-300~\emph{K} (a) Heat capacity of the  TSSG ,TSFZ and polycrystal. The red lines represents  the fitting  of the data according to the Debye-Einstein model for phonon contribution of the heat capacity. The inset represents both the non-phononic heat capacity of the TSSG crystal and the polycrystalline powder extracted by subtracting the phononic contribution(left y -axis) and calculation of the change in entropy of both the samples divided by the value $R*ln2$ (right y-axis) .(b)Heat capacity of polycrystal and TSSG samples at low temperature before and after annealing.}
\label{Fig6}
\end{figure}

\subsection{Heat capacity}
The temperature dependence of the heat capacity of PCTO was investigated for temperature range 0.35-300 \emph{K} on the polycrystalline and TSSG single crystal samples. 


At high temperatures, i.e, above 2~\emph{K} the heat capacity of these two samples is very similar and increases smoothly with temperature indicating the phononic behaviour. Hence, the lattice contribution was fitted according to Debye-Einstein model (DE) for $T_l$ $<$T$<$ 200~\emph{K} using the equation :

\begin{equation}
\begin{split}
C_p(T)&=\left[9(nat-C_{D}-C_{E_i}) R \left(\frac{T}{\theta _D}\right)^{3} \int_{0}^{x_D} {\frac{x^4 {\exp}^x }{(\exp^x -1)^2}} dx\right ]\\
&+\sum_i  C_{E_i}\left[3R(\frac{\theta _E{_i}}{T}) ^ {2} \frac{(\exp\frac{\theta _E{_i}}{T})}{((\exp\frac{\theta _E{_i}}{T})-1)^2}\right]
\end{split}
\label{eq5}
\end{equation}

Where $nat$ is the number of atoms per formula, $C_{D}$, $C_{E}$ are the number of Debye and Einstein modes and $\theta_{D}$, $\theta_{E}$ are corresponding temperatures respectively. The fitting was done using one Debye term and three Einstein terms and $C_{D}$ was fixed to $1$. Here, the lower bound  \emph{$T_l$} was varied between 25-50 \emph{K} in order to accurately extract the magnetic contribution.  Fig.6(a) represents the best fit of the DE for the temperature range 28-195 \emph{K}. The resulting magnetic contribution shows two features for both the samples; a peak  around 1 ~\emph{K} which is sharper for the TSSG sample, and a hump at 11.15($\pm0.04$)~\emph{K} for TSSG and 10.26($\pm0.07$)~\emph{K} for the polycrystal indicating the presence of short-range magnetic correlations(see inset of fig. 6). The peak at 1 ~\emph{K} is discussed in the next section. the error bars were calculated by fitting the hump with Gaussian function.the magnetic entropy analysis on the TSSG single crystal and polycrystalline sample are shown in the inset of fig.6 (a)  in the temperature range 0.35-50 K. The maximum entropy of both TSSG and polycrystal are close to the value expected for a spin-1/2 system at 5.76 J/mol (1 Rln2).  Nevertheless, A detailed entropy analysis on the same samples was reported down to 50 mK in another article \cite{thurn2021}. 



On the other hand, the heat capacity of the polycrystalline and single crystal samples are in stark contrast to each other at low temperatures(see Fig.6(b)). The heat capacity of the polycrystalline sample exhibits a second broad anomaly with maxima $\sim$1.4 \emph{K} and then gradually decreases with temperature. However, the heat capacity of the TSSG single crystal exhibits a sharp lambda-anomaly at 0.98 \emph{K} indicating a phase transition in the system. Also, the broad maximum under the lambda-peak occurs at 1.1 \emph{K}, which is lower than for the poly crystalline sample. Although the lambda-anomaly is mostly suppressed in the powder samples, in some cases  it is observable as a weak peak (see Fig.7).

\subsection{Low temperature transition}

The Lambda anomaly in the heat capacity was reported in the recent paper  by C. Thurn { \it et. al}~\cite{thurn2021}, where it was attributed to a ferroelectric transition accompanied by an anisotropic structural distortion. Here we aim to understand the conditions under which this transition is observed. The difference between polycrystalline and single crystalline samples could be attributed to two possible origins : first, chemical effects like off-stoichiometry, vacancies etc., and second, the sizes of the crystallites in the material. 
\begin{figure}
\includegraphics[width=1 \columnwidth]{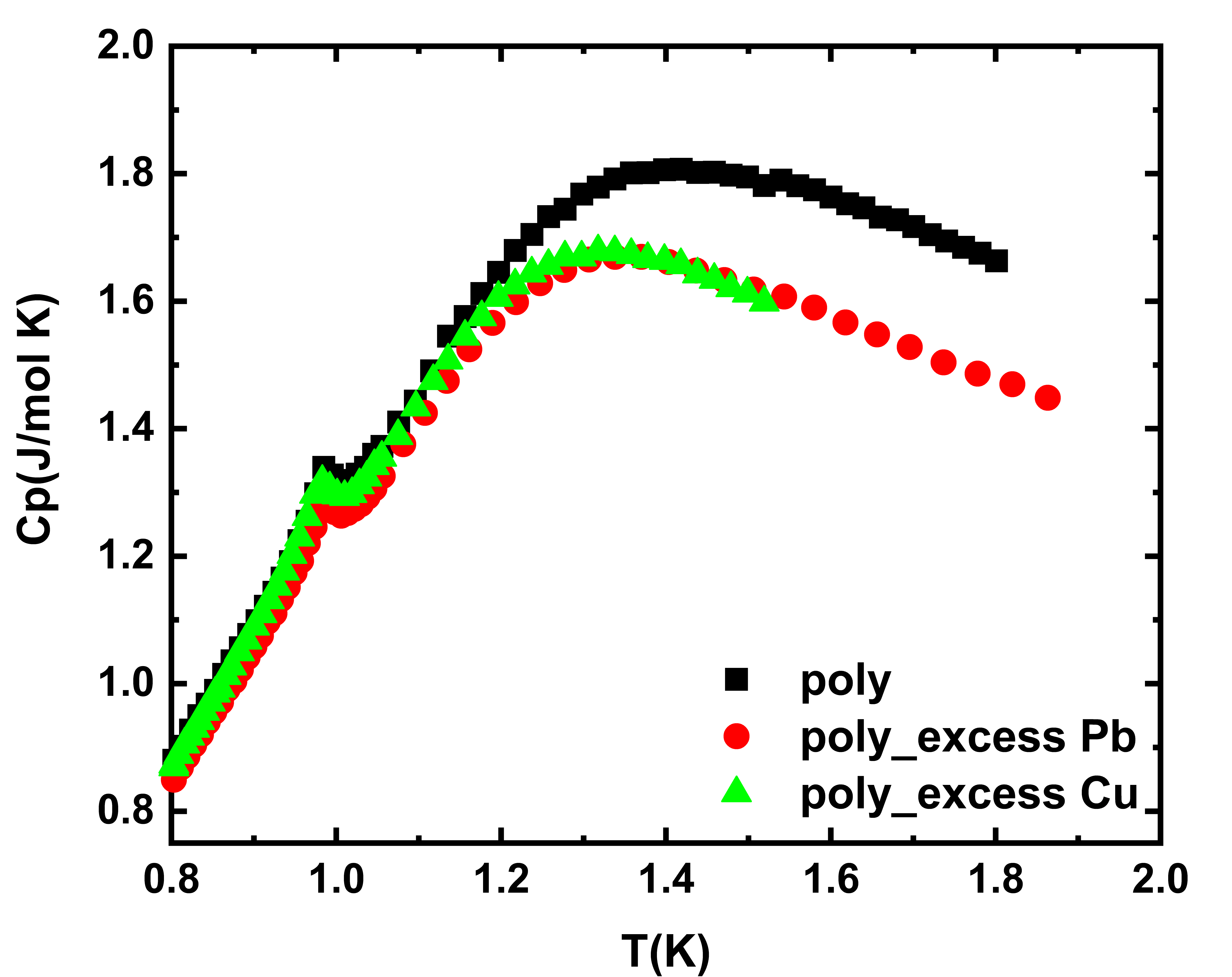}
 
\caption{(Comparison of the heat capacity of \pcto polycrystal with other polycrystalline stoichiometries (2\% excess of Pb and Cu).}
\label{Fig7}
\end{figure}

To test the theory that the discrepancy of the transition at 0.98~\emph{K} between the single crystal and polycrystalline samples of \pcto is due to the possible non-stoichiometric nature of the as-grown powder samples, two additional poly crystalline samples with excess cation ratios are synthesized. The powders which were synthesized following a similar procedure as explained in the experimental part are Pb$_{1.02}$Cu$_{0.98}$Te$_2$O$_6$ and Pb$_{0.98}$Cu$_{1.02}$Te$_2$O$_6$ leading to 2\% excess Pb$^{2+}$ or Cu$^{2+}$ in the system. The resulting powders were characterized by obtaining x-ray diffraction patterns at room temperature. We observe that the nuclear structure of the main phase \pcto and the corresponding intensities of the Bragg peaks in the diffraction pattern are unaffected compared to its stoichiometric counterpart. However, peaks corresponding to PbO and PbTe$_2$O$_5$ appeared in the excess Pb sample (see supplementary information). For the excess Cu sample, extra peaks attributed to CuTe$_2$O$_5$ appeared. 

The heat capacity of these powders, plotted in Fig.7 (with green dots: excess PbO, red dots: excess CuO) reveals that the weak transition anomaly at 0.98~\emph{K} is unaffected suggesting that the stoichiometry of the formed powders is stable i.e, powders do not suffer off-stoichiometry issues with respect to the cation sites and occupancy which was also confirmed by Rietveld refinement of the PXRD. We also found no evidence for Oxygen deficiency in any of the samples. The Oxygen content was measured using TGA where the powders were heated in an Oxygen atmosphere. As a result \pcto is oxidized to PbCuTe$_2$O$_7$ and the change in Oxygen content is determined from the percentage change in mass. The theoretical value of the change would be  2.57 \% if the powder before heating had ideal Oxygen stoichiometry. TGA measurements showed that the experimental values were close to the theoretical value (see supplementary information) and 0.1 percent is the tolerance of the oxygen content in both polycrystalline and TSSG sample.
\begin{figure}
\includegraphics[width=1 \columnwidth]{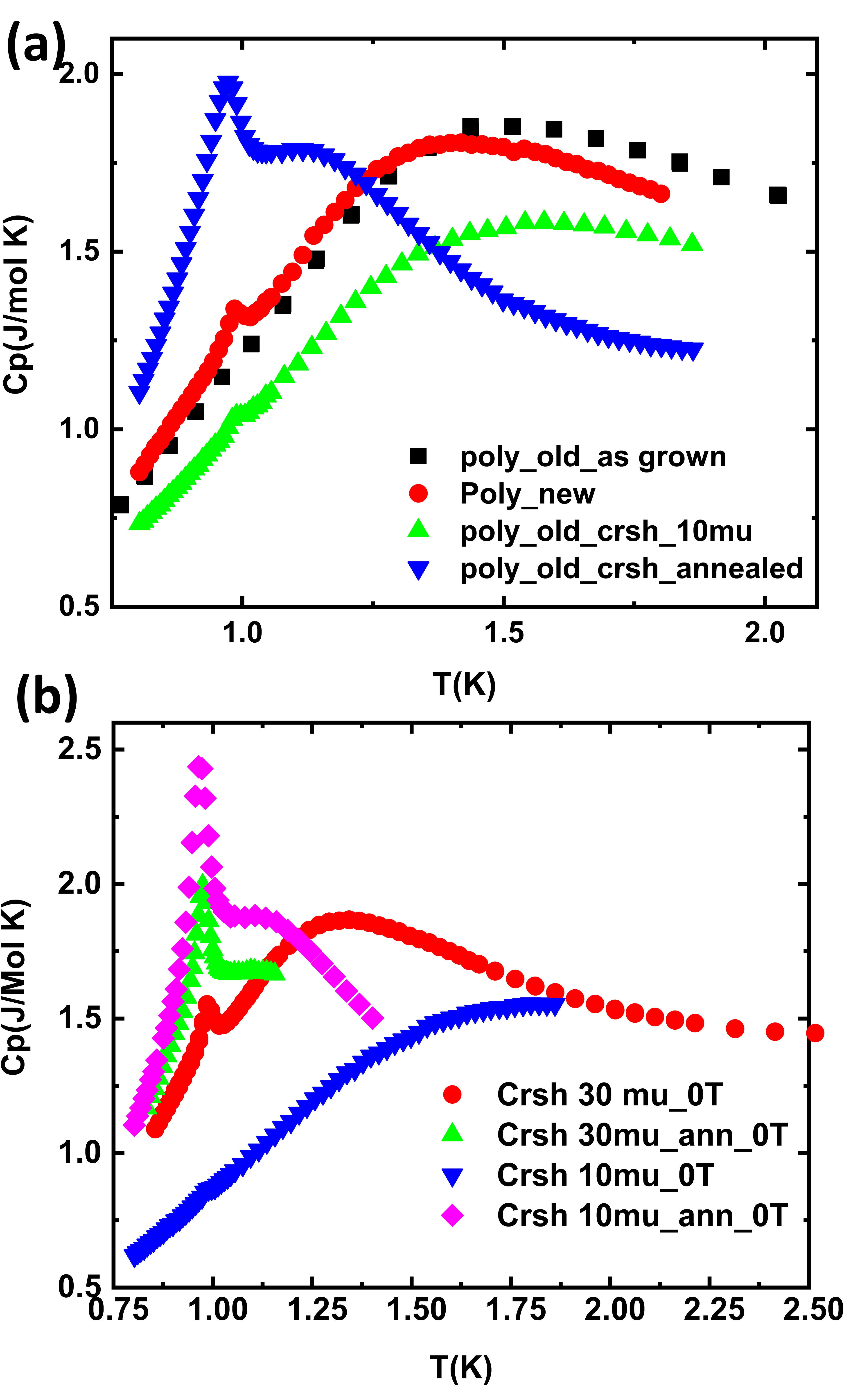}
 
\caption{Size dependence and the effect of annealing on the heat capacity of (a)polycrystal and (b)TSSG samples at low temperature.}
\label{Fig8}
\end{figure}

To investigate the theory that the size of the crystallites is responsible for the discrepancy of the transition at 0.98 \emph{K} between the single crystal and polycrystalline samples of \pcto, changes in the heat capacity were investigated after crushing the samples to reduce the crystallite size and after annealing them to increase the crystallite size and reduce dislocations and grain boundaries. Figure 8(a) shows similar heat capacity for two different batches polycrystalline samples (poly\_old\_as grown, poly\_new) synthesized in a same procedure and with starting grain size approximately 40$\mu$m. In the next step, one of the powders (poly\_old\_as grown) was first crushed to a powder of maximum grain size 10$\mu$m and then subsequently annealed at 525$^\circ$ \emph{C} for 5 days (grain size after annealing 15-40$\mu$m). While the overall heat capacity is reduced by the smaller crystallites due to crushing, annealing reveals features  similar to those observed in the TSSG sample with a sharp peak at 0.98~\emph{K} and the broad underlying peak at 1.1~\emph{K}. This suggests that the improved crystallinity and/or increased crystallite size due to annealing leads to the appearance of the transition in the polycrystalline sample. 

On the other hand, as shown in Fig.8(b), the transition anomaly is weakened drastically upon crushing the single crystal (TSSG) into a fine powder containing crystallites of diameter 30$\mu$m  (red circles). The anomaly almost vanishes when the crystallite size is further reduced (maximum diameter ~10$\mu$m ) and the heat capacity (blue triangles) resembles that of the polycrystalline powder revealing the strong influence of the crystallite size in addition to the crystallinity. This is further confirmed when the transition is recovered to a great extent by annealing these crushed crystals which improves crystallite size as well as the crystallinity within the crystallites(grain size after annealing  26-50$\mu$m). Consequently, the longer annealed powder (magenta diamonds, 5days) exhibits a stronger anomaly compared to the shorter annealing time (green triangles, 2 days). 

The transition  at 1~\emph{K} is inherent to the PCTO system, however, it is strongly influenced by the size and crystallinity of the crystallites. As shown in Fig.6(b), the single crystal shows the sharpest transition while a weaker yet distinctive anomaly is found in the annealed powders which is almost absent in the smaller crystallites with poor crystallinity(Fig.8). A similar scenario has been suggested for a first order structural transition whose width of the hysteresis is inversely proportional to the crystallite size~\cite{Stavrakas2019}. However, the pure structural origin of the transition in the case of \pcto is less likely due to the low temperatures involved. On the other hand, the dielectric anomaly indicating the ferroelectric transition also appears only in the samples with bigger crystallite size (i.e., single crystals)~\cite{thurn2021}. Therefore it is plausible that the transition is in fact a second order ferroelastic transition accompanied by a structural transition below which cubic structure breaks into lower symmetry ( eg: tetragonal or orthorhombic) twin domains. In this case, both the ferroelctric as well as structural anomaly at 0.98~\emph{K} are absent when the  crystallite size is smaller than a critical domain size as has been observed in prototypical ferroelctric materials~\cite{Ishikawa1988,Arlt1990,Chattopadhyay1995}.

\section*{Summary and conclusions}
In summary, single crystals of \pcto were grown using both top-seeded solution growth (TSSG)  and travelling solvent optical floating zone (TSFZ) techniques. The challenges in obtaining a pure phase were solved by analyzing the phase diagram and thermal properties of the compound. The difficulty faced in the TSFZ technique to maintain a stable melt-zone for a prolonged time could be bypassed using the TSSG technique and higher quality, large, and impurity-free single crystals could  be grown. Both the single crystals and the polycrystalline samples were analyzed using powder x-ray diffraction, magnetic, and heat capacity measurements. Single crystals revealed an anomaly in the heat capacity  around at 0.98~\emph{K} which is nearly absent or much smaller in the polycrystalline samples. The crystallite size of the powders in \pcto play a strong role in the appearance of the transition. The sharp $\lambda$-peak which was previously shown to be due to ferroelectric order~\cite{thurn2021} is accompanied by a structural transition in PCTO below which domains are formed due to twinning as a result of symmetry lowering e.g. from the cubic to tetragonal or cubic to orthorhombic. The transition is intrinsic to PCTO and its suppression in fine-powder samples is unrelated to differences in sample quality or impurities, but is rather due to a higher energy cost of ferroelectric domain formation within small crystallites.

The magnetic properties of PCTO could be affected by the structural transition which would probably alter the bond lengths of the first and second nearest neighbour Cu$^{2+}$-Cu$^{2+}$ distances making the corresponding interaction strengths different from each other, and maybe also splitting each interaction into several inequivalent interactions. This scenario is expected to alter the degree frustration in the low temperature phase. In case the frustration is lowered allowing magnetic order to occur, the expected ordered moment will be very small as the spectral weight of the excitations is lowered below $\sim$0.4~meV at lower temperature~\cite{Chillal2020}. When compared to the simulated diffraction magnetic Bragg peak and the energy scale of the spin waves in the isostructural unfrustrated compound SrCuTe$_2$O$_6$ which develops long-range magnetic order ($1.25$ \emph{meV})~\cite{Chillal20212,Chillal20213}, the ordered moment in PCTO can be estimated to be less than 0.1$\mu${$_B$}. It should however be noted that the magnetic measurement of  C.Thurn {\it et.al}~\cite{thurn2021} do not find an accompanying magnetic transition, rather suggesting that the highly frustrated state survives in the single crystal below the transition with enhanced quantum critical behavior compared to the powders. Therefore, further measurements focusing on the low temperature crystal structure and magnetic properties are required to determinine the ground state of \pcto. 

\section*{Acknowledgements}
We thank Ch. Thurn, P. Eibisch, A. Ata, U. Tutsch, Y. Saito, S. Hartmann, J. Zimmermann, B. Wolf and M. Lang ( Goethe-Universität Frankfurt(M) ) who first discover the \emph{T = 1 K} transition in \pcto and who supported us with helpful insights into the origins of this feature. This work is supported by the Deutsche Forschungsgemeinschaft (DFG) through the project B06 of the SFB-1143 (ID:247310070). The powder synthesis, crystal growth, and physical properties measurements took place at the Core Laboratory Quantum Materials, Helmholtz Zentrum Berlin für Materialien und Energie, Germany.

\bibliographystyle{apsrev4-2}
\bibliography{pcto_bib}{}

\end{document}